# Influence of Gd-rich precipitates on the martensitic transformation, magnetocaloric effect and mechanical properties of Ni-Mn-In Heusler alloys – A comparative study

Franziska Scheibel[1*], Wei Liu[1], Lukas Pfeuffer[1], Navid Shayanfar[1], Andreas Taubel[1], Konstantin P. Skokov[1], Stefan Riegg[1], Yuye Wu[1,2], Oliver Gutfleisch[1]

[1]Technical University of Darmstadt, Institute of Materials Science, Alarich-Weiss-Straße 16, 64287 Darmstadt, Germany
[2]Key Laboratory of Aerospace Materials and Performance (Ministry of Education), School of Materials Science and Engineering, Beihang University, Beijing 100191, People's Republic of China

*franziska.scheibel@tu-darmstadt.de

Keywords: metal matrix composites, magnetocaloric, magneto-structural phase transitions, microstructure, mechanical properties, Heusler alloy

## Abstract

A multi-stimuli cooling cycle can be used to increase the cyclic caloric performance of multicaloric materials like Ni-Mn-In Heusler alloys. However, the use of a uniaxial compressive stress as an additional external stimulus to a magnetic field requires good mechanical stability. Improvement of mechanical stability and strength by doping has been shown in several studies. However, doping is always accompanied by grain refinement and a change in transition temperature. This raises the question of the extent to which mechanical strength is related to grain refinement, transition temperature, or precipitates. This study shows a direct comparison between a single-phase Ni-Mn-In and a two-phase Gd-doped Ni-Mn-In alloy with the same transition temperature and grain size. It is shown that the excellent magnetocaloric properties of the Ni-Mn-In matrix are maintained with doping. The isothermal entropy change and adiabatic temperature change are reduced by only 15% in the two-phase Ni-Mn-In-Heusler alloy compared to the single-phase alloy, which is resulting from a slight increase in thermal hysteresis and the width of the transition. Due to the same grain size and transition temperature, this effect can be directly related to the precipitates. The introduction of Gd precipitates leads to a 100% improvement in mechanical strength, which is significantly lower than the improvement observed for Ni-Mn-In alloys with grain refinement and Gd precipitates. This reveals that a significant contribution to the improved mechanical stability in Gd-doped Heusler alloys is related to grain refinement.

## Introduction

Ni-(Co)-Mn-X (X: In, Sn, Sb, Ga, Ti) Heusler alloys have been receiving large attention during the last years due to their magnetic shape-memory and magnetocaloric properties, which makes these materials potential candidates for magnetic refrigeration [1–4]. Like other magnetocaloric materials exhibiting a first-order magneto-structural phase transformation (FOMST), also Ni-Mn-In is suffering from the drawback of the intrinsic thermal hysteresis which limits the magnetocaloric effect (MCE) under cyclic conditions [4–7]. Besides the MCE, also large barocaloric or elastocaloric effects are observed in Ni-Mn-In Heusler alloys [8–11]. The possibility to induce the FOMST by two kinds of external stimuli makes Ni-Mn-In an excellent material for the multi-stimuli refrigeration cycle described by Gottschall et al. [12]. These multi-stimuli refrigeration cycle turns the thermal hysteresis of first-order caloric materials into an advantage. The hysteresis is here used to prevent a reverse transformation after inducing the FOMST by a magnetic field. Since the removal of the magnetic field does not induce a reverse transformation, the exposure time and the volume of the permanent magnet can be significantly reduced [12,13]. The reverse transformation is afterwards induced by a second stimulus, namely uniaxial stress.

The multi-stimuli refrigeration cycle requires functional materials exhibiting both excellent magnetocaloric and elastocaloric as well as mechanical properties. In contrast to the magnetocaloric cycle (single stimulus), the increase of thermal hysteresis is even beneficial for the multi-stimuli cycle to prevent the reverse transformation. The second stimulus is then used to overcome the hysteresis. Ni-Mn-In shows excellent caloric properties like a sharp FOMST, large shift of the FOMST with magnetic field and uniaxial stress as well as a large isothermal entropy change $\Delta s_T$ and adiabatic temperature change $\Delta T_{ad}$ [3,11,14,15]. However, the material is quite brittle, which limits the mechanical cyclic stability for multi-stimuli cooling [16,17]. The multi-caloric material must withstand not only the internal stress due to the volume change during the FOMST, but also the external applied uniaxial stress. A promising strategy to improve the mechanical stability of Ni-Mn-X (X: Ga, In, Sn) Heusler alloys, besides grain refinement [15], is doping [17–26]. The enhanced mechanical stability by doping goes hand in hand with grain refinement [17,21,24–27] and/or the formation of precipitates [18–23,26]. Especially, the addition of rare earth (RE) elements, such as Gd, [21,24], Dy[28], Tb [27,29,30], Y[31] leads to the formation of RE-rich precipitates since the RE elements are not soluble in the Ni-Mn-X (X: Ga, In) Heusler phase. The compressive strength and strain of this Heusler alloys can be increased up to 2-3 times by RE doping [24,29]. However, the doping or substitution of the element X in Ni(Co)Mn-X (X: Ga, In) with RE-elements (Gd, Tb, Y) changes the chemical composition of the Ni-Mn-X matrix phase, which also changes drastically the temperature of the FOMST [24,28,29,31]. This makes it difficult to compare the caloric properties of the RE-doped and single-phase Ni-Mn-X alloys since the caloric properties (thermal hysteresis, shift of the FOMST, $\Delta s_T$ and $\Delta T_{ad}$) vary depending on the chemical composition and temperature of the FOMST [32,33]. A direct comparison of the caloric properties between a doped and a non-doped Ni-Mn-X alloy requires the same temperature of FOMST for both samples, this is investigated in this study.

The improvement of mechanical properties by doping is validated by several studies [21,24,26,29,31]. However, in these studies, doping causes simultaneous changes in several characteristic properties that can be considered for improving mechanical properties: Grain size refinement, formation of precipitates and change in the chemical composition of the matrix phase (change in transition temperature). There is to the best of our knowledge no comparison between doped and non-doped Ni-Mn-based Heusler alloys with comparable grain size, equal FOMST temperature and chemical composition of the Ni-Mn-X matrix phase, this allows as novel method to directly extract the influence of the precipitates on the hysteresis in this study.

Our study compares the magnetic, and magnetocaloric properties ($\Delta s_T$ and $\Delta T_{ad}$) of a Gd-doped Ni-Mn-In alloy with a Ni-Mn-In reference alloy exhibiting comparable FOMST temperature and grain size. The Gd-doped Ni-Mn-In alloy contains a RE-rich secondary phase surrounded by a Ni-Mn-In-matrix. The matrix and the reference sample have nearly the same chemical composition, leading to a FOMST temperature in close proximity. By keeping the transition temperature, chemical composition and grain size equal, changes in the magnetocaloric performance and mechanical properties can be directly linked to the presence of RE-rich precipitates. It is shown that in Gd-doped Ni-Mn-In the good magnetocaloric properties could be preserved while at the same time the mechanical stability could be improved. However, the mechanical stability is far lower than in alloys combining both Gd-doping and grain refinement[21,34], which proves that the grain-refinement plays a major role.

## Material and methods

A nominal $Ni_{50}Mn_{35}In_{15}$ composition was prepared by arc melting of high purity Ni (99.97%), Mn (99.9%) and In (99.99%) from *chemPUR*. Due to the evaporation of Mn during the melting an excess of 3 at% was added. The sample was subsequently annealed in a quartz tube under Ar atmosphere at 1073 K for 90 h, followed by rapid quenching in water. This sample serves as the reference without doping. For the Gd-doped sample 1 at.% Gd was added to the nominally composed $Ni_{50}Mn_{35}In_{15}$. Melting and annealing procedure are the same as for the reference sample. In the text, the samples will be denoted as $Ni_{50}Mn_{35}In_{15}$ and $(Ni_{50}Mn_{35}In_{15})$+Gd.

The actual composition was determined by energy-dispersive X-ray spectroscopy (EDX) using a *EDAX Octane Plus* detector and a *Tescan Vega 3* scanning electron microscope (SEM). The microstructure and distribution of Gd-rich precipitates were determined by backscatter electron (BSE) imaging as well as by optical microscopy using a *Zeiss Axio Imager.D2m* with polarized light function. High-resolution SEM images are determined using a JEOL 7600 microscope.

X-ray powder diffraction (XRD) has been used to determine the crystallographic structure. The measurements were done using a Stoe Stadi P diffractometer with Mo $K_{\alpha 1}$ radiation, in transmission mode in a $2\theta$ range of 10° to 50°. The Rietveld refinements were performed using FullProf/WinPLOTR suite software [35,36]

For the compression tests, the samples were cut into 2.5 mm × 2.5 mm × 5 mm cubes. The measurements were conducted by an *Instron 5967* universal testing machine with a maximum force of 30 kN using a constant displacement rate of $5 \times 10^{-3}$ mms$^{-1}$.

The magnetic characterization was performed with a *LakeShore 7410* vibrating sample magnetometer (VSM) and a Quantum Design physical property measurement system (PPMS-14 T). Isofield curves of magnetization were determined with a cooling and heating rate of 2 Kmin$^{-1}$. Isothermal curves of magnetization were determined with a field-application rate of 5 mTs$^{-1}$. A discontinuous temperature protocol was performed before each measurement to ensure a defined initial state and exclude effects of transformation history. For this, the sample was heated to the full austenite and cooled to the full martensite state before the measurement temperature was set. For the isothermal minor hysteresis loops no discontinuous temperature protocol was used and the sample remains at the measured temperature. The isothermal entropy changes during the first-order transition were calculated by using the Maxwell relation

$$\Delta s_T(T, \Delta H) = \int_{H_1}^{H_2} \left( \frac{\partial M(T,H)}{\partial T} \right)_H dH . \quad (1)$$

$\partial M(T,H)/\partial T$ was determined by temperature-dependent magnetization measurements under isofield condition in fields from 0.25 to 2 T with equidistant field-steps of 0.25 T.

Differential scanning calorimetry (DSC) measurements were performed in the temperature range 170 K ≤ $T$ ≤ 370 K using a liquid nitrogen cooled setup *Netzsch 404 F1* (Silver furnace). To increase the sensitivity of the sample-holder (Type E thermocouple) a He/Ar gas flow of 70 ml/min and aluminum crucibles were used. The sensitivity and temperature calibration for the applied heating rate 5 Kmin$^{-1}$ were carried out in advance using reference materials and a baseline with two empty crucibles to correct the setup influence.

The adiabatic temperature change $\Delta T_{ad}$ was measured directly in a specifically developed device generating a sinusoidal field-sweep profile with a maximum field of 1.93 T. $\Delta T_{ad}$ was measured by a differential type T thermocouple. A detailed description can be found in Ref. [3]. The measurements are performed using the discontinuous protocol, in which the sample is first heated above the austenite finish temperature ($A_F$) to ensure pure austenitic phase (270 K), then the sample is cooled below the martensite finish temperature ($M_F$) to ensure a pure martensitic phase (230 K) [32]. After that initial procedure, the sample is heated up to the measurement temperature to determine $\Delta T_{ad}$.

## Results and discussion

### 1.1 Structural and microstructural characterization

The microstructure of the reference sample ($Ni_{50}Mn_{35}In_{15}$) and the Gd-doped sample ($Ni_{50}Mn_{35}In_{15}+Gd$) is investigated using optical microscopy (Fig. 1 (a) and (c)), SEM (Fig. 1 (c) and (d)) and high-resolution SEM (Fig. 1 (e) and (f)) covering the different length scale from mm to μm range. The reference sample (a) exhibits a coarse columnar grain structure with grain sizes up to several millimeters in length and up to 300 µm in width. This microstructure is typical for arc-molten Ni-Mn-In ingots, where the grains grow along the solidification direction (perpendicular to the contact surface with the copper base plate) [15,37]. The $(Ni_{50}Mn_{35}In_{15})$+Gd sample (c) shows the same coarse columnar grain structure as the Gd-free $Ni_{50}Mn_{35}In_{15}$ sample. A grain refinement by RE-doping, as reported for Ni-Co-Mn-In-Gd [24] or arc molten Ni-Mn-Ga-Tb [25,38], cannot be observed here. A larger magnification of the surface of the two samples is shown in Fig. (b) and (d). Fig. 1(d) shows the presence of the Gd-rich secondary phase in the form of spherical precipitates (bright contrast) with a size of several micrometers. The precipitates are equally distributed within the whole sample and the individual grains. The brighter contrast of the precipitates relates to the material contrast between the matrix and the secondary phase (precipitates). For comparison, the Gd-free $Ni_{50}Mn_{35}In_{15}$ sample (b) shows no presence of secondary phases. The dark spots relate to small pores, polishing related artifacts and impurities of manganese oxides. High-resolution SEM images of the Ni-Co-Mn-In-Gd (Fig. (e) and (f)) show that the Gd-rich precipitates have a round or oval shape with a diameter between 1 and 5µm.

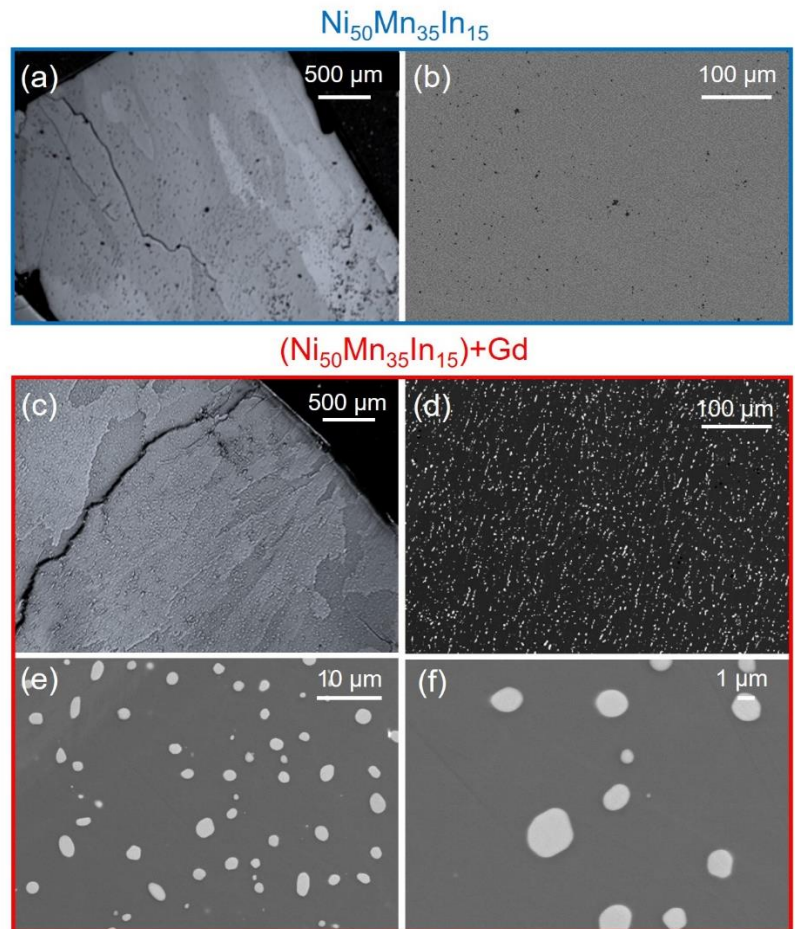

Fig. 1: (a) (c) Optical microscopy and (b) (d) BSE images of $Ni_{50}Mn_{35}In_{15}$ and $(Ni_{50}Mn_{35}In_{15})$+Gd. (e)(f) High-resolution SEM images of $(Ni_{50}Mn_{35}In_{15})$+Gd show the Gd-rich precipitates (bright phase) in the Ni-Mn-In matrix.

An elemental mapping of the $(Ni_{50}Mn_{35}In_{15})$+Gd sample is shown in Fig. 2. Fig. 2(a) shows the BSE image of the Gd-rich phase surrounded by a Ni-Mn-In matrix. An EDX elemental mapping of the selected area (blue rectangle) is shown in Fig. 2(b). Mapping confirms that Gd is present only in the secondary phase, the Mn content in the precipitates is much lower than in the matrix, whereas the In concentration is only slightly lower. The content of Ni is similar in both phases. Point spectra of different precipitates and areas are performed to verify the composition of the Gd-rich precipitates and the Ni-Mn-In matrix, the results are shown in Tab. 1 together with the composition of the $Ni_{50}Mn_{35}In_{15}$ reference alloy. The composition of the $(Ni_{50}Mn_{35}In_{15})$+Gd matrix is nearly identical to the one of reference $Ni_{50}Mn_{35}In_{15}$ sample leading to nearly the same electron per atom ratio within the range of the measurement error. Since the first-order transition of Ni-Mn-based Heusler alloys are strongly correlated to the e/a ratio [2,39], the matrix should yield a similar transition temperature as the reference sample, which will be shown in the following. A clear determination of the precipitate composition cannot be done by EDX analysis due to the large deviation of the different point spectra. The deviation can be explained by the size of the precipitates (5-10 µm) which is similar to the excitation pear of the electron beam and therefore the spectra contain element information of both precipitate and matrix. However, the small deviation of the In content indicates that both phases have an In content about 14 to 15 at.%. Considering the results of Li *et al.*[24], indexing a hexagonal $CaCu_5$ type structure for the Gd-rich precipitates in NiCoMnInGd, the precipitates can be assumed to be a $GdNi_5In$ or $GdNi_4In$ phase [31,40,41]. From acquired BSE images, a surface fraction of 5% Gd-rich precipitates and 95% Ni-Mn-In matrix could be determined by binary contrast analysis.

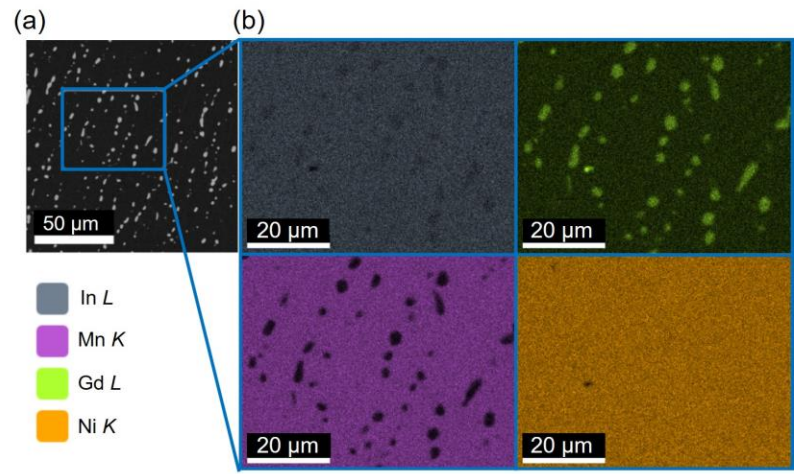

Fig. 2: (a) BSE image of $(Ni_{50}Mn_{35}In_{15})$+Gd, the bright contrast corresponds to the Gd-rich precipitates surrounded by the Ni-Mn-In matrix. (b) EDX mapping of the selected area (marked in blue in (a)), for the elements In, Mn, Gd and Ni.

Tab. 1: Average values for the composition and electron per atom ratio e/a of $(Ni_{50}Mn_{35}In_{15})$+Gd (matrix and precipitates) and $Ni_{50}Mn_{35}In_{15}$ determined by EDX point analysis, the standard deviation is given as error.

| Sample | Ni [at.%] | Mn [at.%] | In [at.%] | Gd [at.%] | e/a |
|---|---|---|---|---|---|
| $(Ni_{50}Mn_{35}In_{15})$+Gd matrix | 47.7 ± 0.3 | 37.7 ± 0.2 | 14.6 ± 0.2 | -- | 7.85± 0.05 |
| $(Ni_{50}Mn_{35}In_{15})$+Gd precipitates | 51.7 ± 4.6 | 27.4 ± 11.5 | 15.0 ± 0.6 | 5.9 ± 6.3 | |
| $Ni_{50}Mn_{35}In_{15}$ | 48.2 ± 0.6 | 36.9 ± 0.1 | 14.9 ± 0.5 | -- | 7.86± 0.09 |

Fig. 3(a) depicts the X-ray diffraction pattern for the $Ni_{50}Mn_{35}In_{15}$ reference sample measured at room temperature. The Rietveld refinement confirms the $L2_1$ structure of the austenite phase with a lattice constant of $a = 0.60053(1)$ nm. The diffraction pattern of the $(Ni_{50}Mn_{35}In_{15})$+Gd sample (Fig. 3(b)) shows additional minor peaks indicated by *. These minor peaks correspond to the Gd-rich precipitates and can be indexed by a hexagonal $CaCu_5$ type structure. However, a clear indexing is not possible due to the low phase fraction. Furthermore, the ternary system Gd-Ni-In also reveals a large variety of different possible composition with different crystal structures [42]. Therefore, it is also possible that the Gd-rich precipitates vary in composition and structure. For more details, the angular range around the minor peaks, marked in turquoise is shown in (c), using a logarithmic scale. The Ni-Mn-In matrix of $(Ni_{50}Mn_{35}In_{15})$+Gd is in the austenite state showing a $L2_1$ structure with a lattice constant of $a = 0.60108(3)$ nm. This lattice constant, as well as the lattice constant of the reference sample are in agreement with the literature of comparable stoichiometry [33,43]. The slight variation of the lattice constants of the two samples can be explained by the slightly different chemical composition, see Tab. 1. The diffraction pattern of the $(Ni_{50}Mn_{35}In_{15})$+Gd sample indicates a broadening of the peaks related to the Ni-Mn-In matrix. The broadening of the peak is related to the Lorentzian part of the peak shape which can affected by grain size or strain effects. The overall grain size of both samples is comparable, as it is visible in Fig.1. As the sample is quenched in water, the difference in the thermal expansion coefficient in the Ni-Mn-In and Gd-rich phase causes an increase of residual stress. The lattice mismatch of the matrix and the Gd-rich precipitates can also cause a residual stress [44,45], with respect to the μm-size of the precipitates and the large inter-precipitate distance this is considered a minor effect. However, in the interface region of the Gd-rich precipitates and the Ni-Mn-In matrix, chemical inhomogeneities in the Ni-Mn-In matrix can be present and lead to a broadening of the peaks.

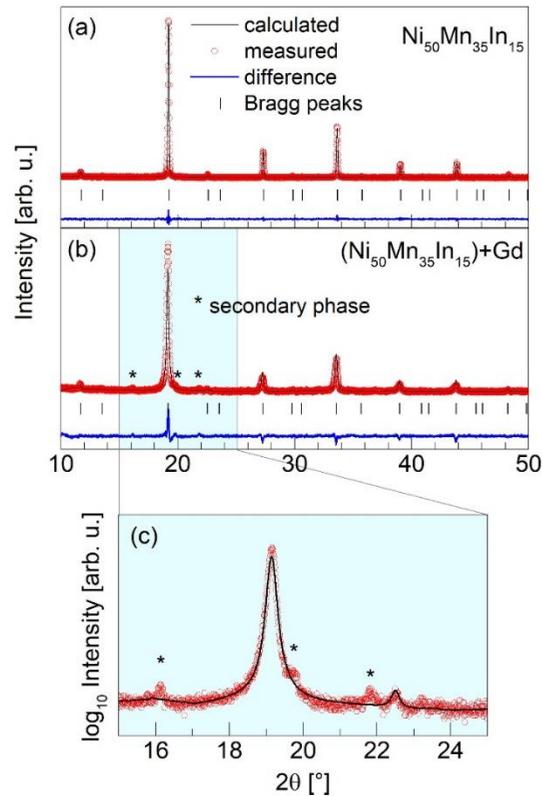

Fig. 3: X-ray diffraction patterns of (a) $Ni_{50}Mn_{35}In_{15}$ and (b) $(Ni_{50}Mn_{35}In_{15})$+Gd at room temperature. Rietveld refinement confirms the $L2_1$ structure of the austenite state, the diffraction peaks indicated by * are related to the Gd-rich precipitates. (c) shows the area marked in turquoise with logarithmic intensity.

*1.2 Mechanical behavior*

Fig. 4 depicts the compressive stress-strain curves for $(Ni_{50}Mn_{35}In_{15})$+Gd and $Ni_{50}Mn_{35}In_{15}$. The stress-strain curve of the polycrystalline $Ni_{50}Mn_{35}In_{15}$ samples can be divided into two regions AB and BC. Between A and B the sample shows a linear behavior related to the elastic deformation. With increasing stress, the $Ni_{50}Mn_{35}In_{15}$ undergoes a plastic deformation indicated by a decrease in slope (BC) until the sample breaks at C. In a single phase Heusler alloy, cracking occurs at the grain boundary. Due to the large grain size of this sample, this leads to a complete fracture of the sample before an austenite-to-martensite transformation can be induced. The elastic deformation of the Gd-doped $(Ni_{50}Mn_{35}In_{15})$+Gd sample is comparable to the single-phase alloy. At about 200 MPa the $(Ni_{50}Mn_{35}In_{15})$+Gd undergoes an austenite-to-martensite transformation $(\sigma_{Ms})$. The critical stress $(\sigma_{Ms})$ observed for $(Ni_{50}Mn_{35}In_{15})$+Gd (Fig. 4) fits to the stress field-driven shift of the transition temperature of 0.24 KMPa$^{-1}$ reported in literature [14,15].

The multiple small dips in the stress-strain curve between $\sigma_{Ms}$ and C indicate the formation of microcracks during the transformation. The Gd-rich precipitates seem to act as barriers for the transformation shearing, leading to the formation of microcracks as the transformation progresses. The complete fracture of the sample occurs at C once sufficient microcracks have been accumulated. The strength of the $(Ni_{50}Mn_{35}In_{15})$+Gd alloy is greatly increased compared to the single phase alloy. It can be concluded that the presence of the Gd-rich precipitates within the Ni-Mn-In grains has a strengthening effect on the alloy and stops the crack propagation during transformation. Since the grain size and chemical composition of the matrix phase of both samples are comparable, the strengthening effect is directly related to the presence of the Gd-rich precipitates, grain refinement or compositional variation can be excluded. The degree of strengthening resulting from the precipitation formation is rather small which can be explained by the large size [46] and the spherical shape [47] of the precipitates (Fig. 1(e) (f)). With a size up to 10 µm a precipitation strengthening by the Orowan dislocation bypassing mechanism is most likely expected [46]. Since the yield strength increment decrease with increasing precipitate size, the µm-size Gd-rich precipitates lead only to a small strengthening of the alloy. For a more coherent precipitation strengthening the precipitate size should be reduced to the nm-range [46]. Compared to the Gd-doped Ni-Mn-In alloy shown in Fig. 4, a much higher strengthening of RE-doped Ni-Mn-based alloys is observed in Ref.[21,24,27,38] . The precipitate size in these studies is comparable to the one presented here, but the grain size in these studies is significantly decreased by doping. This shows that the major effect of the strengthening is the grain refinement whereas the precipitation strengthening plays a minor role.

However, the formation of microcracks during stress-induced transformation can lead to a premature failure of the alloy after several cycles. A detailed analysis of the microcracks formation and propagation under cyclic condition is subject of current investigation.

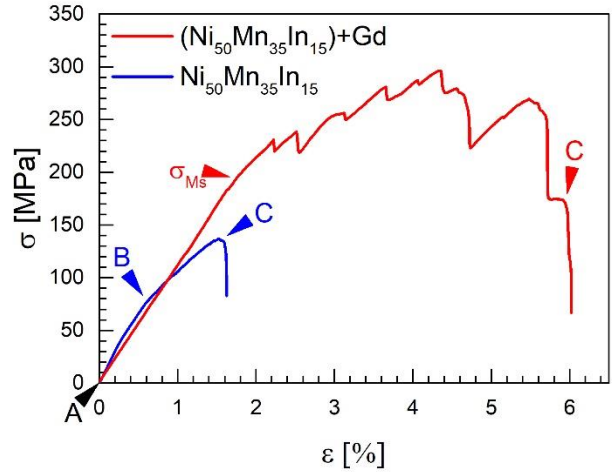

Fig. 4: Compressive stress-strain curves of $Ni_{50}Mn_{35}In_{15}$ and $(Ni_{50}Mn_{35}In_{15})+Gd$, the measurements are performed at room-temperature using a constant displacement rate of $5 \cdot 10^{-3} mms^{-1}$.

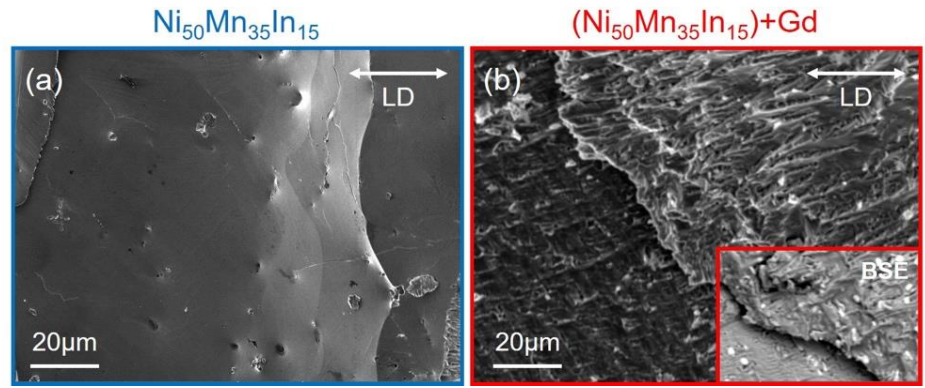

Fig. 5: SE images of the fracture surface of (a) $Ni_{50}Mn_{35}In_{15}$ and (b) $(Ni_{50}Mn_{35}In_{15})+Gd$. The smaller inset in (b) shows the BSE image of the corresponding surface, the bright sports correspond to the Gd-rich precipitates. The loading direction (LD) is indicated by arrows.

To explain the mechanical properties in more detail, the fracture surface of both samples was analyzed. Fig. 5 shows the SE image of the fracture surface of $Ni_{50}Mn_{35}In_{15}$ and $(Ni_{50}Mn_{35}In_{15})+Gd$, the loading direction (LD) is indicated by arrows. The $Ni_{50}Mn_{35}In_{15}$ alloy (Fig. 5(a)) shows a smooth surface indicating an intergranular fracture along the grain boundary which is typical for brittle materials like Ni-Mn-based Heusler alloys. In comparison, a transgranular fracture is observed in the $(Ni_{50}Mn_{35}In_{15})+Gd$ alloy (Fig. 5(b)), which matches the higher compressive strength and strain. The small inset in Fig. 5(b) shows the BSE image of the corresponding fracture surface, the Gd-rich secondary phase corresponds to the bright phase. As also shown in Figs. 1 and 2, the secondary phase is homogeneously distributed within the Ni-Mn-In matrix, which hinders the crack-growth and changes the fracture mechanism towards a transgranular fracture. Such a change is also observed in other RE-doped Ni-Mn-based Heusler alloys [25,31].

*1.3 Martensite-Austenite transformation behavior*

The temperature-dependent magnetization of (Ni$_{50}$Mn$_{35}$In$_{15}$)+Gd and Ni$_{50}$Mn$_{35}$In$_{15}$ in an applied field of 1 T are shown in Fig. 6. Both samples show a first-order magnetostructural transition from low temperature low magnetic martensite to high temperature high magnetic austenite in the temperature range between 190 and 255 K and exhibit nearly the same start temperature for the austenite-to-martensite phase transformation ($M_S$). The transformation is completed at $M_F$ which is lower in the (Ni$_{50}$Mn$_{35}$In$_{15}$)+Gd sample. For the reverse transformation (martensite-to-austenite) a higher start and finishing temperature ($A_S$ and $A_F$) is observed for (Ni$_{50}$Mn$_{35}$In$_{15}$)+Gd compared to Ni$_{50}$Mn$_{35}$In$_{15}$. The thermal hysteresis is 14.5 K for Ni$_{50}$Mn$_{35}$In$_{15}$ and 20.0 K for (Ni$_{50}$Mn$_{35}$In$_{15}$)+Gd. All values are listed in Tab. 2. The comparison both magnetization curved blow and above the transition region (190 K > $T$ > 255 K) reveal a slightly lower magnetization curve for the (Ni$_{50}$Mn$_{35}$In$_{15}$)+Gd sample. The effect is more dominant in the temperature range above the transition where the Ni-Mn-In matrix is in the ferromagnetic state. This relates to the presence of PM precipitates [48] which are reducing the amount of Ni-Mn-In matrix in the sample and therefore the net magnetization of the sample. The increase of the thermal hysteresis can be caused by a variety of intrinsic and extrinsic parameters like, magnetic and chemical ordering, grain size, defects, phase purity or residual and interface stress [7,44]. Since the chemical composition of the (Ni$_{50}$Mn$_{35}$In$_{15}$)+Gd matrix and Ni$_{50}$Mn$_{35}$In$_{15}$ reference is nearly the same, as it is for the grain size and transition temperature, the increase of the thermal hysteresis can be directly linked to the presence of the Gd-rich precipitates. As shown in Fig. 3 the precipitates increase the residual stress which causes the increase in the thermal hysteresis.

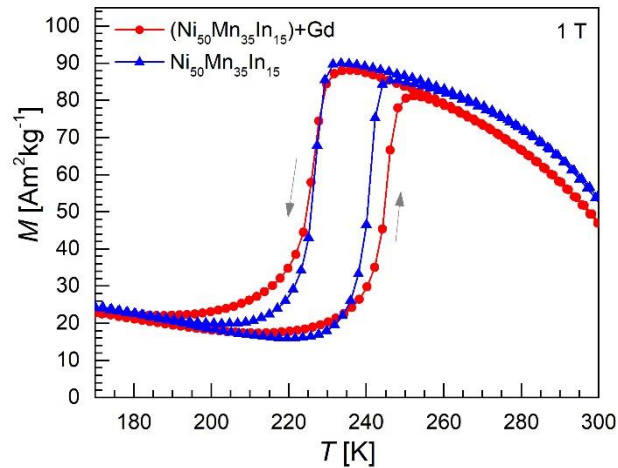

*Fig. 6: Temperature-dependent magnetization of (Ni$_{50}$Mn$_{35}$In$_{15}$)+Gd and Ni$_{50}$Mn$_{35}$In$_{15}$ at 1 T. The field cooling and heating curves are indicated by arrows.*

*Tab. 2: Start ($M_S$, $A_S$) and finish ($M_F$, $A_F$) temperatures of the martensite-to-austenite phase transition of (Ni$_{50}$Mn$_{35}$In$_{15}$)+Gd and Ni$_{50}$Mn$_{35}$In$_{15}$. The table also contain the thermal hysteresis $\Delta T_{hyst}$ and Curie-temperature $T_C$ of both alloys, all data are determined in a field of 1T.*

| Sample | $A_S$ [K] | $A_F$ [K] | $M_S$ [K] | $M_F$ [K] | $\Delta T_{hyst}$ [K] | $T_C$ [K] |
|---|---|---|---|---|---|---|
| (Ni$_{50}$Mn$_{35}$In$_{15}$)+Gd | 239 | 249 | 231 | 220 | 20.0 | 311.8 |
| Ni$_{50}$Mn$_{35}$In$_{15}$ | 236 | 244 | 230 | 222 | 14.5 | 315.5 |

The shift of the transition temperature with applied magnetic field is an important parameter for the magnetocaloric effect. Fig. 7 shows the isofield, temperature-dependent magnetization measurements for (a) $Ni_{50}Mn_{35}In_{15}$ and (b) $(Ni_{50}Mn_{35}In_{15})$+Gd in applied fields from 0.25 to 2 T with equidistant field-steps of 0.25 T. For sake of clarity, only the curves at 0.5, 1.0, 1.5 and 2.0 T are shown. Both samples show a linear decrease of the transition temperature with increasing external field. The transition width and thermal hysteresis remain constant within this field range. Fig. 7(c) shows the transition temperatures $T_t$ ($M_S$, $M_F$, $A_S$ and $A_F$) for the isofield $M(T)$ measurements, determined by linear approximation of the transition regions [32]. A linear fitting of the transition temperatures shows a constant shift $dT_t/\mu_0 dH$ of -6.8 ($A_S$), -5.2 ($A_F$), -5.7 ($M_S$) and -8.2 $KT^{-1}$ ($M_F$) for $Ni_{50}Mn_{35}In_{15}$ and -6.2 ($A_S$) -5.8($A_F$), -6.8 ($M_S$) and -6.4 $KT^{-1}$ ($M_F$) for $(Ni_{50}Mn_{35}In_{15})$+Gd. $dT_t/\mu_0 dH$ of $Ni_{50}Mn_{35}In_{15}$ and $(Ni_{50}Mn_{35}In_{15})$+Gd are comparable with literature values [32,49]. Therefore, we can conclude that the presence of the Gd-rich precipitates does not affect $dT_t/\mu_0 dH$ of the Ni-Mn-In matrix.

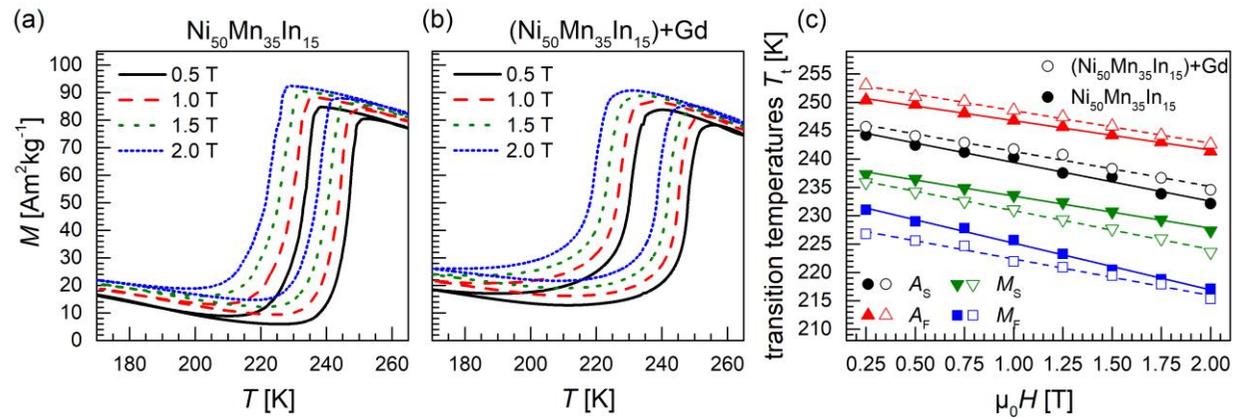

Fig. 7: Temperature-dependent magnetization of (a) $Ni_{50}Mn_{35}In_{15}$ and (b) $(Ni_{50}Mn_{35}In_{15})$+Gd measured in magnetic fields of 0.5, 1.0, 1.5 and 2.0 T. (c) Magnetic field dependence of the martensite-austenite transition temperatures $M_S$, $M_F$, $A_S$, $A_F$ of $(Ni_{50}Mn_{35}In_{15})$+Gd (full) and $Ni_{50}Mn_{35}In_{15}$ (open). The temperatures are determined by the tangent method [32]. A linear regression is used to determine $dT_t/\mu_0 dH$.

To investigate the field-induced transformation from martensite to austenite state in more detail, field-dependent magnetization measurements are performed. Fig. 8 shows field-dependent magnetization of $Ni_{50}Mn_{35}In_{15}$ and $(Ni_{50}Mn_{35}In_{15})$+Gd in fields up to 10 T at temperatures below (100 K), above (280 K) and in the vicinity of the FOMST (180-240 K). At 100 K both samples are in the martensitic phase and show a week ferromagnetic behavior. For the temperatures between 180 K and 220 K the magnetic field-induces a fully transformation to the austenite phase, the transformation is reversed during demagnetization. Thereby, the critical temperatures for the start and the finish of the transformation decreases with increasing temperature. The field-induced transformation of $Ni_{50}Mn_{35}In_{15}$ and $(Ni_{50}Mn_{35}In_{15})$+Gd differs only with respect to $\Delta T_{hyst}$, where $(Ni_{50}Mn_{35}In_{15})$+Gd has a larger $\Delta T_{hyst}$. This behavior is consistent with the results of the temperature-induced transformation in Fig. 7. At 240 K both samples undergo a transformation to the austenite phase, however the reverse transformation cannot be induced during demagnetization due to $\Delta T_{hyst}$ and both samples remain in the austenite start after the field is removed. The decrease of the magnetization below 1 T corresponds to the demagnetization of the FM austenite phase. At 280 K no transformation is induced since both samples are in the austenite phase.

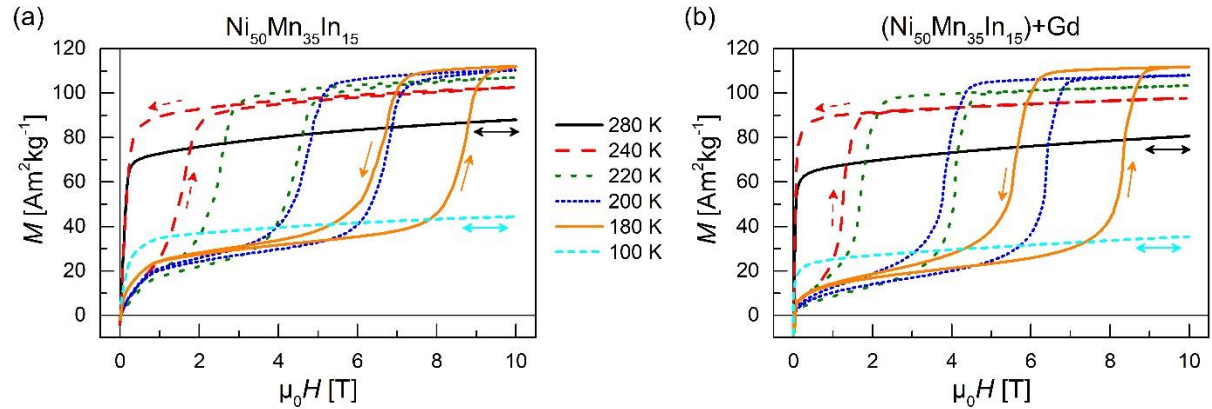

Fig. 8: Field-dependent magnetization of (a) $Ni_{50}Mn_{35}In_{15}$ and (b) $(Ni_{50}Mn_{35}In_{15})$+Gd measured at different temperatures between 100 K (martensite) and 280 K (austenite). For the temperatures between 180 K and 240 K a transformation from martensite to austenite is induced by the applied magnetic field.

## 1.4 Caloric effect and the effect of hysteresis

### Field induced transition and entropy change

The magnetocaloric effect depends strongly on the magnetization change during FOMST and on $dT_t/\mu_0 dH$. Since both parameters are comparable for $Ni_{50}Mn_{35}In_{15}$ and $(Ni_{50}Mn_{35}In_{15})$+Gd we assume that also the magnetocaloric effects are similar. One magnetocaloric quantity is the isothermal entropy change $\Delta s_T$. We calculated $\Delta s_T$ by the Maxwell relation (1) using the isofield temperature-dependent magnetization measurements shown in Fig. 7(a) and (b). Fig. 9 shows the temperature-dependent $\Delta s_T$ for $Ni_{50}Mn_{35}In_{15}$ and $(Ni_{50}Mn_{35}In_{15})$+Gd for an applied field of 1 and 2 T. The $(Ni_{50}Mn_{35}In_{15})$+Gd alloy shows a maximum $\Delta s_T$ of 9.2 and 7.0 $Jkg^{-1}K^{-1}$ in fields of 2 and 1 T, respectively. The maximum $\Delta s_T$ of the single-phase $Ni_{50}Mn_{35}In_{15}$ alloy are 10.7 and 7.6 $Jkg^{-1}K^{-1}$ in fields of 2 and 1 T, respectively. The Gd-rich precipitates therefore reduce the maximum $\Delta s_T$ by 14% and 8% in fields of 2 and 1 T, respectively. The maximum $\Delta s_T$ of both samples fit to the reported values in literature between 9.8 and 11.8 $Jkg^{-1}K^{-1}$ [32,49]. The peak-like shape of all $\Delta s_T(T)$ curves indicate that the transformation to the austenite phase is not completely induced by an applied field up to 2 T. The full transformation can be induced by applying higher fields [13,50]. However, even the partial transformation of $(Ni_{50}Mn_{35}In_{15})$+Gd shows a $\Delta s_T$ value comparable to the one in single-phase $Ni_{50}Mn_{35}In_{15}$. The slightly lower maximum $\Delta s_T$ of 14% and 8% in fields of 2 and 1 T for the $(Ni_{50}Mn_{35}In_{15})$+Gd alloy is related to the slight broadening of the FOMST due to Gd-doping. A larger width ($A_S$-$A_F$) of the transition reduces the phase fraction transformed from the martensite into the austenite phase by the applied magnetic field which leads to a reduced $\Delta s_T$.

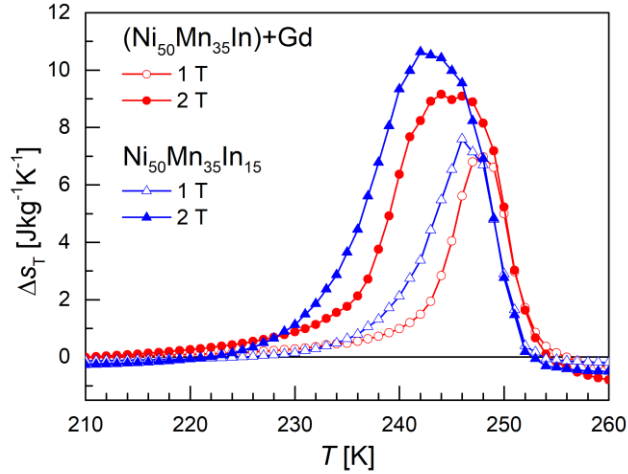

*Thermal induced transition and entropy change*

To compare $\Delta s_T$ for a complete FOMST, DSC measurements were performed. Fig. 10 shows the temperature dependent DSC signal for (a) Ni$_{50}$Mn$_{35}$In$_{15}$ and (b) (Ni$_{50}$Mn$_{35}$In$_{15}$)+Gd under heating, using a rate of 5 Kmin$^{-1}$. The transition temperature $A_S$ and $T_t$ as well as $T_C$ agree with the temperatures determined by the temperature-dependent magnetization. $\Delta s_T$ is determined by the area underneath the endothermic peak at $T_t$ (gray area). A $\Delta s_T$ of 11 Jkg$^{-1}$K$^{-1}$ and 13 Jkg$^{-1}$K$^{-1}$ are determined for Ni$_{50}$Mn$_{35}$In$_{15}$ and (Ni$_{50}$Mn$_{35}$In$_{15}$)+Gd, respectively. Considering a conservative measurement error of 10% for the DSC measurements, no reduction of $\Delta s_T$ for a complete FOMST can be observed in Gd-doped Ni$_{50}$Mn$_{35}$In$_{15}$ using a Gd-content of 1 at.%.

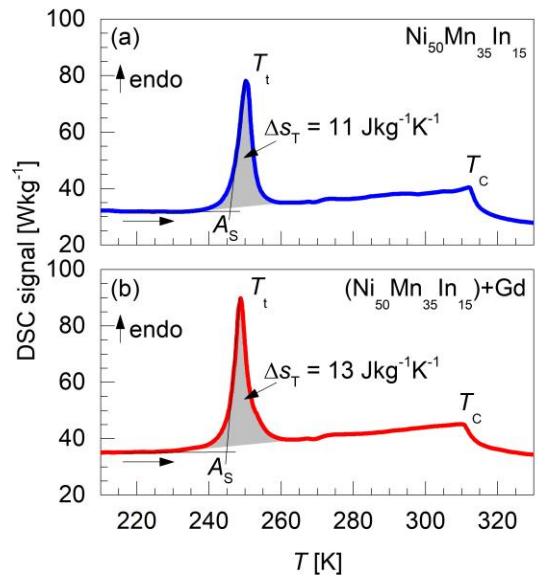

*Fig. 10: Differential scanning calorimetry (DSC) heating curves of (a) Ni$_{50}$Mn$_{35}$In$_{15}$ and (b) (Ni$_{50}$Mn$_{35}$In$_{15}$)+Gd using a heating rate of 5 Kmin$^{-1}$. The transition temperatures $A_S$, $T_t$ and $T_C$ are indicated, the gray area underneath the endothermic peak corresponds to the entropy change $\Delta s_T$ of the FOMST.*

*Effect of hysteresis on the thermoelastic phase transition*

The hysteresis of the transition is caused by irreversible or dissipative energy losses $E_{irr}^{M-P}$ in the transformation process, and leads to a reduced magnetocaloric effect under cyclically applied magnetic field [5,7,51]. This dissipation of energy therefore additionally affects the efficiency of the magnetocaloric material in the magnetocaloric refrigeration cycle, since less magnetic work can be converted into refrigeration work. Regarding the thermodynamic model for thermoelastic martensitic transitions from Orin and Planes [52], the phase transformation from martensite to parent phase (austenite) (*M-P*) take place under the condition of local equilibrium of the Gibbs free energy change $\Delta G^{M-P}$. The equilibrium can be written as the balance between the chemical and elastic contribution of the free energy change and the irreversible energy losses.

$$\Delta G^{M-P} = \Delta G_{ch}^{M-P} - \Delta G_{el}^{M-P} + E_{irr}^{M-P} = 0$$

The chemical contribution $\Delta G_{ch}^{M-P}$ arises from the difference in the Gibbs free energy between austenite and martensite and acts as the driving force of the transformation and depends on the temperature and the magnetic field. The reversible elastic strain energy $\Delta G_{el}^{M-P}$ originates from interfacial energy at single or multiple interfaces (including interfacial energy at the habit-plane of the martensite and austenite phase, grain boundaries, and precipitates[53]), and the elastic strain energy due to elastic strains [52]. Shamberger and Ohuchi [51] estimated $\Delta G_{el}^{M-P}$ and $E_{irr}^{M-P}$ as

$$\Delta G_{el}^{M-P} = \frac{1}{2}\Delta s_t \Delta T_{el} \qquad \text{with} \qquad \Delta T_{el} = [(A_F - A_S) + (M_S - M_F)]/2$$
$$E_{irr}^{M-P} = \Delta s_t \Delta T_{irr} \qquad \text{with} \qquad \Delta T_{irr} = [(A_F + A_S) - (M_S + M_F)]/2$$

where $\Delta T_{el}$ is correlated to the transition width, and $\Delta T_{irr}$ is the average thermal hysteresis of the material. The increase in the transition width and the thermal hysteresis between doped and non-doped Ni-Mn-In has been detected in Fig. 6, indicating an increase of $\Delta G_{el}^{M-P}$ and $E_{irr}^{M-P}$. The elastic contribution and the dissipative energy losses of $Ni_{50}Mn_{35}In_{15}$ and $(Ni_{50}Mn_{35}In_{15})$+Gd are calculated using the transition temperatures at zero field and $\Delta s_T$ determined from DSC measurements (complete FOMST at zero field). The transition temperatures at zero field are determined by the extrapolation of the linear fit in Fig. 7(c), all values are listed in Tab. 3.

*Tab. 3: Calculated $G_{el}^{M-P}$ and $E_{irr}^{M-P}$ for $Ni_{50}Mn_{35}In_{15}$ and $(Ni_{50}Mn_{35}In_{15})$+Gd, using the zero-field transitions temperatures and $\Delta s_T$ determined by DSC measurements. The values of the zero-field transitions temperatures are determined by the extrapolation of the linear fit in Fig. 7(c).*

|  | $Ni_{50}Mn_{35}In_{15}$ | $(Ni_{50}Mn_{35}In_{15})$+Gd |
|---|---|---|
| $A_S$ ($\mu_0 H$=0) [K] | 246.2 | 247.5 |
| $A_F$ ($\mu_0 H$=0) [K] | 252.0 | 254.3 |
| $M_S$ ($\mu_0 H$=0) [K] | 239.1 | 237.7 |
| $M_F$ ($\mu_0 H$=0) [K] | 233.4 | 228.8 |
| $\Delta s_T$ (DSC) [Jkg$^{-1}$K$^{-1}$] | 11±1 | 13±1 |
| $E_{irr}^{M-P}$ [Jkg$^{-1}$] | 141±13 | 229±17 |
| $\Delta G_{el}^{M-P}$ [Jkg$^{-1}$] | 32±3 | 51±4 |

Comparing the Gd-doped and the reference sample, an increase of $E_{irr}^{M-P}$ of 62% can be observed for the sample with Gd-rich precipitates. Regarding $\Delta G_{el}^{M-P}$, an increase of 61% can be determined for the sample with Gd-rich precipitates compared to the reference sample containing no Gd. The presence of the Gd-rich precipitates increases the dissipative energy losses and the number of interfaces and therefore also increases the interfacial energy. In addition, the precipitates can lead to an increase of local strain in the Ni-Mn-In matrix in the vicinity of the precipitate. Since both samples exhibit the same grain size, an increase of the dissipative energy losses and the interfacial energy due to grain refinement can be excluded.

The thermoelastic model to determine $\Delta G_{el}^{M-P}$ and $E_{irr}^{M-P}$ is an approximation and does not include an asymmetric hysteresis or transformation behavior for forward and reverse transformation [51,52], or an increase of the dissipative energy by increasing applied magnetic field [54,55]. Therefore, the calculation of $E_{irr}^{M-P}$ includes a number of uncertainties, however, since this model is applied to both samples, exhibiting a nearly identical chemical composition of the matrix phase and grain size, the relative difference of $\Delta G_{el}^{M-P}$ and $E_{irr}^{M-P}$ between both samples can be determined and directly linked to the presence of the Gd-precipitates.

*Field induced adiabatic temperature change*

The second important figure of merit for magnetocaloric compounds is the adiabatic temperature change $\Delta T_{ad}$. Fig. 11 shows the temperature-dependent $\Delta T_{ad}$ for $Ni_{50}Mn_{35}In_{15}$ and $(Ni_{50}Mn_{35}In_{15})$+Gd for a field change of 1.9 T. The full symbols in Fig.7 show the temperature-dependent $\Delta T_{ad}$ at the first application of 1.9 T. After removing the magnetic field, the field was applied for a second time (open symbols). Due to the thermal hysteresis the $\Delta T_{ad}$ under cyclic conditions is reduced. In the vicinity of the FOMST (around 250 K) a maximum $|\Delta T_{ad}|$ of 4.8 K and 4.3 K is measured for $Ni_{50}Mn_{35}In_{15}$ and $(Ni_{50}Mn_{35}In_{15})$+Gd for the first field application, respectively. The values are comparable with literature values of Ni-Mn-In Heusler alloys with similar chemical composition [32]. Since the $Ni_{50}Mn_{35}In_{15}$ sample and the Ni-Mn-In matrix of the $(Ni_{50}Mn_{35}In_{15})$+Gd sample have the nearly the same e/a ratio and $T_C$, a similar MCE at $T_C$ is expected. This is confirmed by the maximum conventional $\Delta T_{ad}$ of 2.0 K and 1.9 K for $Ni_{50}Mn_{35}In_{15}$ and $(Ni_{50}Mn_{35}In_{15})$+Gd, respectively. The slightly lower $\Delta T_{ad}$ of $(Ni_{50}Mn_{35}In_{15})$+Gd can be explained by the slightly broader FOMST, which is also the reason for the reduced $\Delta s_T$ in Fig. 9. A good comparability of the samples can be achieved by using higher magnetic fields as this would induce a complete transformation, which is part of our further research. The thermal hysteresis does not affect the $|\Delta T_{ad}|$ of the first application cycle but affects the reversible MCE measured in the second application cycle (open symbols). Here both samples show a reduced maximum $|\Delta T_{ad}|$ of 0.7 K for $Ni_{50}Mn_{35}In_{15}$ and 0.6 K for $(Ni_{50}Mn_{35}In_{15})$+Gd compared to the first application cycle. The difference in the thermal hysteresis for $Ni_{50}Mn_{35}In_{15}$ and $(Ni_{50}Mn_{35}In_{15})$+Gd is 5.5 K, which explains also the slightly lower $|\Delta T_{ad}|$ of $(Ni_{50}Mn_{35}In_{15})$+Gd for the second application cycle. However, in a multi-stimuli cooling cycle the thermal hysteresis can be overcome by applying the second stimulus and the $\Delta T_{ad}$ of the first field application can be fully repeated under cyclic conditions [12]. The comparison of maximum $|\Delta T_{ad}|$ of the reference and Gd-doped samples reveals a reduction by 10% relate to the presence of Gd-rich precipitates.

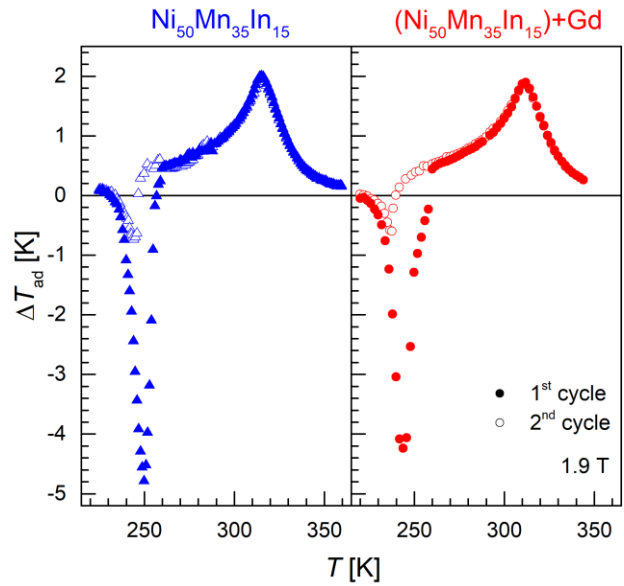

## Summary and conclusion:

The formation of Gd-rich precipitates in multicaloric Ni-Mn-In improves the mechanical stability while the good magnetocaloric properties are maintained. This is an essential development for the usage of the material in a multi-stimuli cooling cycle. The spherical precipitates with a size between 1 and 10 µm are homogeneously distributed within the Ni-Mn-In matrix. In contrast to casting into 6 mm diameter bars [24], the preparation by arc melting and subsequent annealing at 1073 K for 90 h shows no grain refinement in combination with Gd-doping. By comparing Gd-doped $Ni_{50}Mn_{35}In_{15}$ alloy with a single-phase reference alloy with nearly equal chemical composition and grain size, the improvement in the mechanical stability in this study can directly be linked to the presence of Gd-rich precipitates, since grain refinement is not present. The analysis of the fracture surface indicates a change of fracture mechanism towards transgranular fracture by Gd-doping. The Gd-rich precipitates act as barriers for the transformation shearing which leads to the formation of microcracks during the transformation. The FOMST temperature, the transition temperature shift $dT_t/\mu_0 dH$, the saturation magnetization and $T_C$ of the Ni-Mn-In matrix are not affected by the Gd-rich precipitates. The main effect of the Gd-rich precipitates can be observed in terms of the thermal hysteresis and the width of the transition. Both parameters are increased in the ($Ni_{50}Mn_{35}In_{15}$)+Gd sample with Gd-rich precipitates compared to the reference alloy without precipitates. This leads to an increase in the irreversible or dissipative energy losses and the reversible elastic strain energy of the FOMST by about 61% compared to the Gd-free reference sample. These parameters are also affecting the magnetocaloric properties. A $\Delta s_T$ of 9.2 $Jkg^{-1}K^{-1}$ and a $|\Delta T_{ad}|$ of 4.3 K for a magnetic-field change of 2 T have been measured for the Gd-doped Ni-Mn-In alloy. These values are only slightly decreased by 10% for $|\Delta T_{ad}|$ and 14% for $\Delta s_T$ compared to the values determined for the reference sample. Therefore, Gd-doping of Ni-Mn-based Heusler can be used to improve the mechanical stability. However, even better mechanical properties are achieved by a combination of Gd-doping and grain refinement [21,34]. This study shows that the effect of grain refinement plays a major role for the improvement of the mechanical stability.

## Acknowledgements:

This work is supported by funding from the European Research Council (ERC) under the European Union's Horizon 2020 research and innovation program (grant no. 743116 - Cool Innov) and from the DFG (SPP 1599 and CRC/TRR 270 "HoMMage" Project-ID 405553726-TRR 270). The authors thank Fernando Maccari for the HR-SEM.

## Data availability statement:

The data that support the findings of this study are available from the corresponding author upon reasonable request.

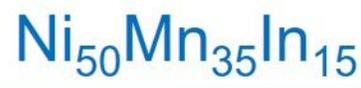

$Ni_{50}Mn_{35}In_{15}$

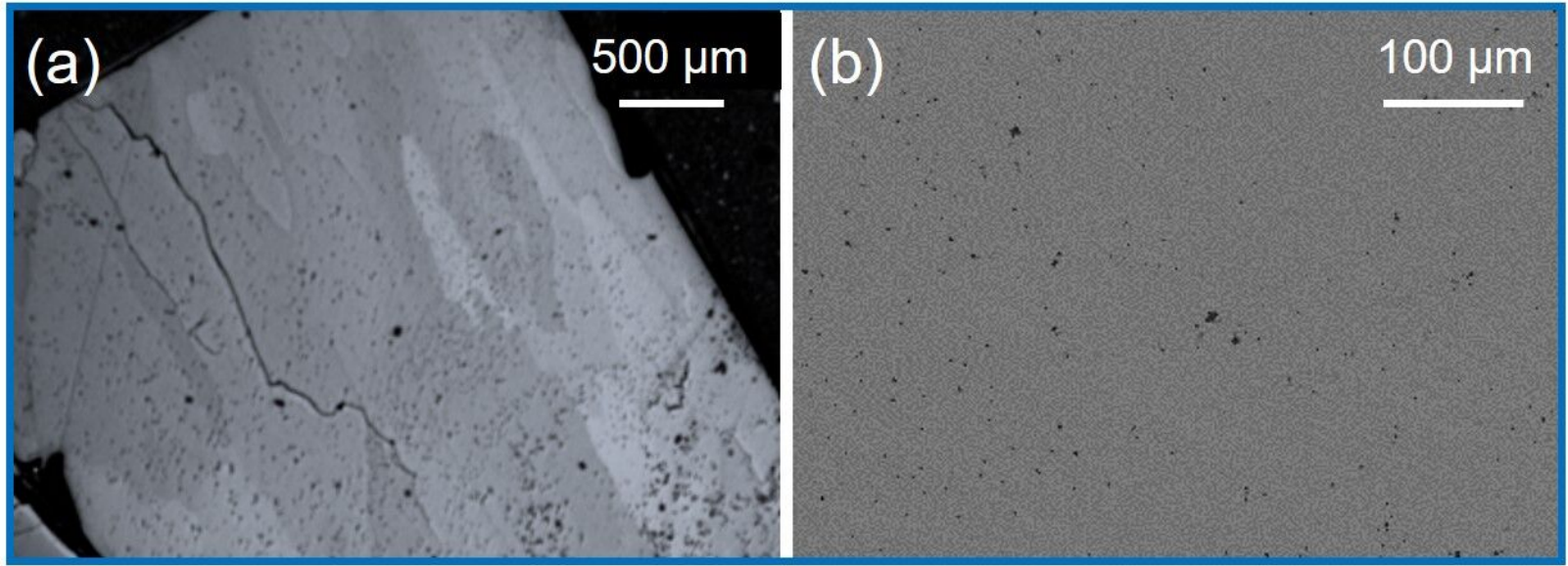

$(Ni_{50}Mn_{35}In_{15})+Gd$

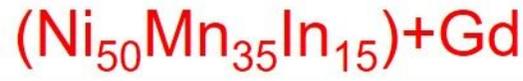

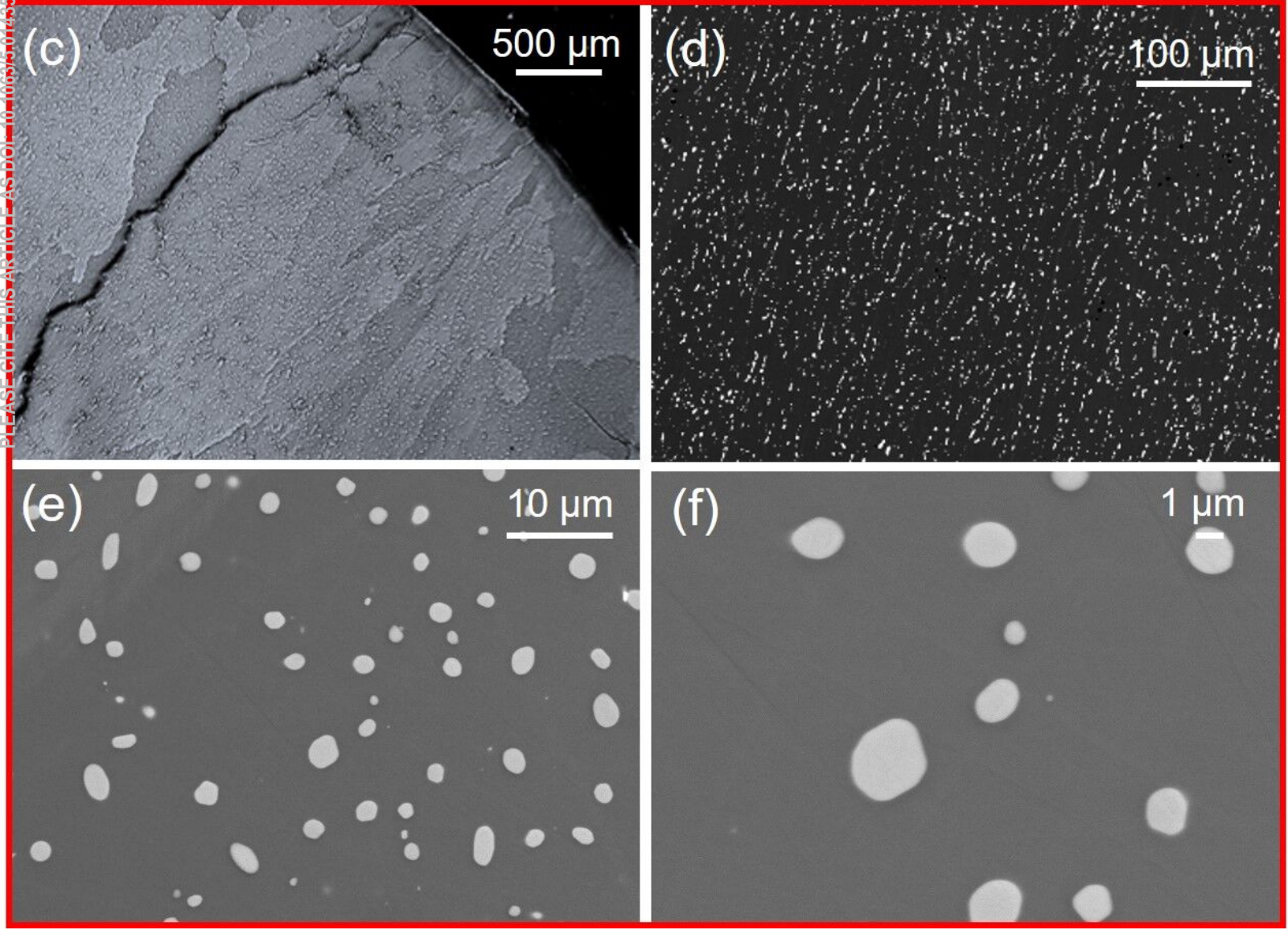

Journal of
Applied Physics

(a)

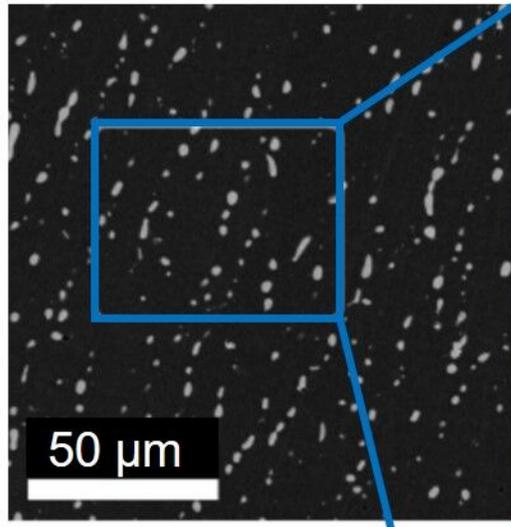

(b)

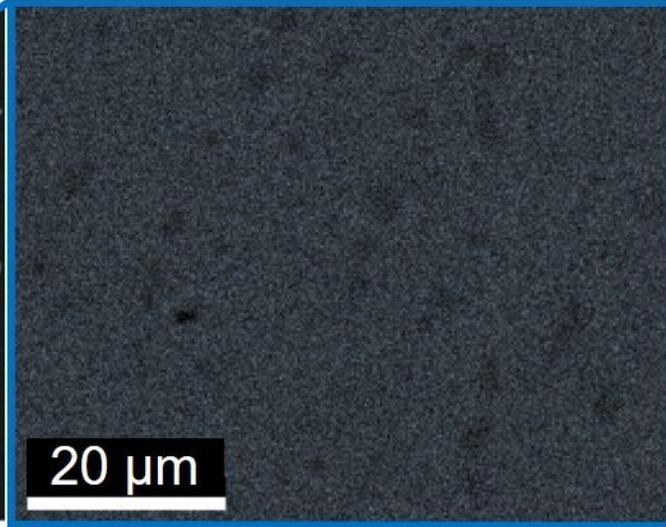
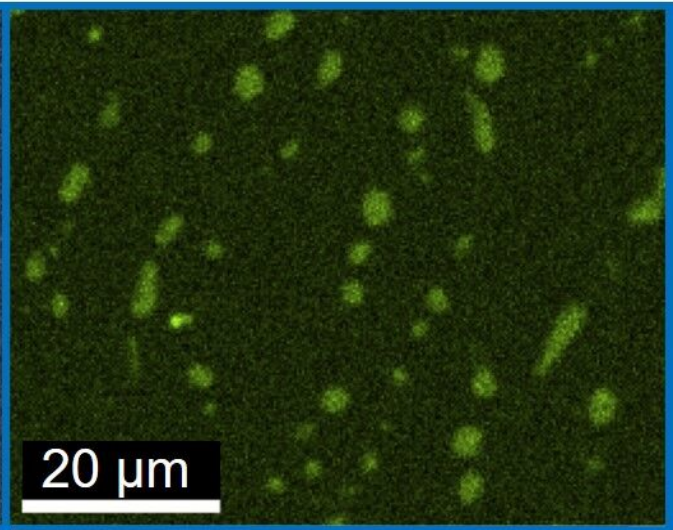

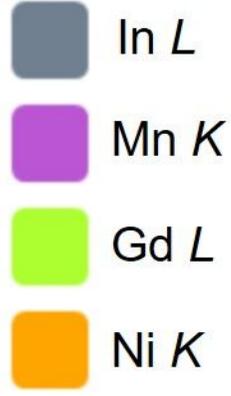

In *L*

Mn *K*

Gd *L*

Ni *K*

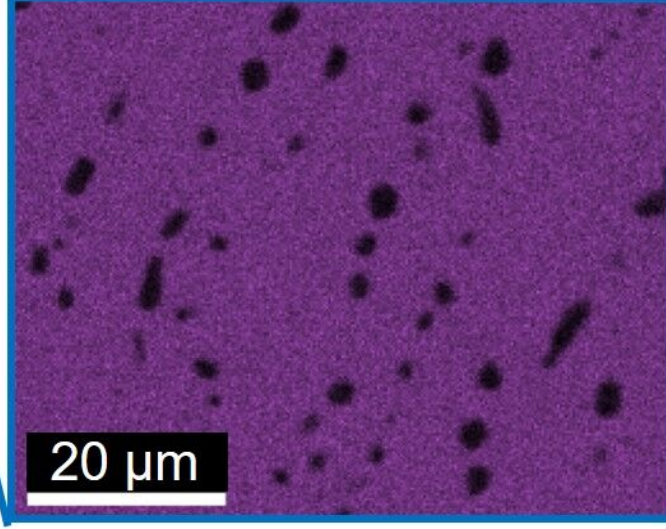
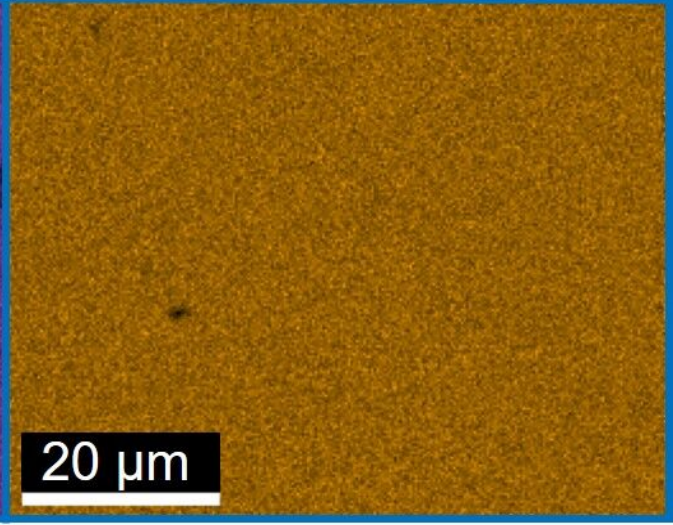

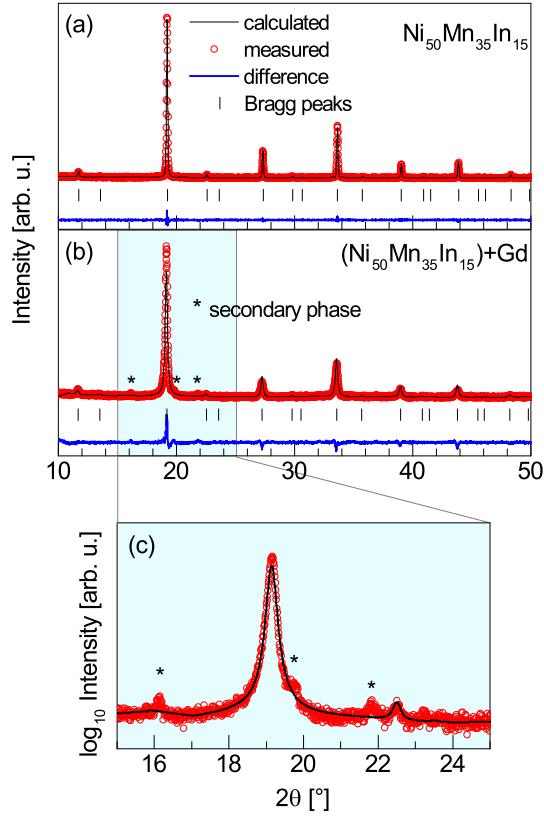

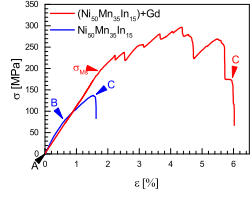

Ni$_{50}$Mn$_{35}$In$_{15}$

(Ni$_{50}$Mn$_{35}$In$_{15}$)+Gd

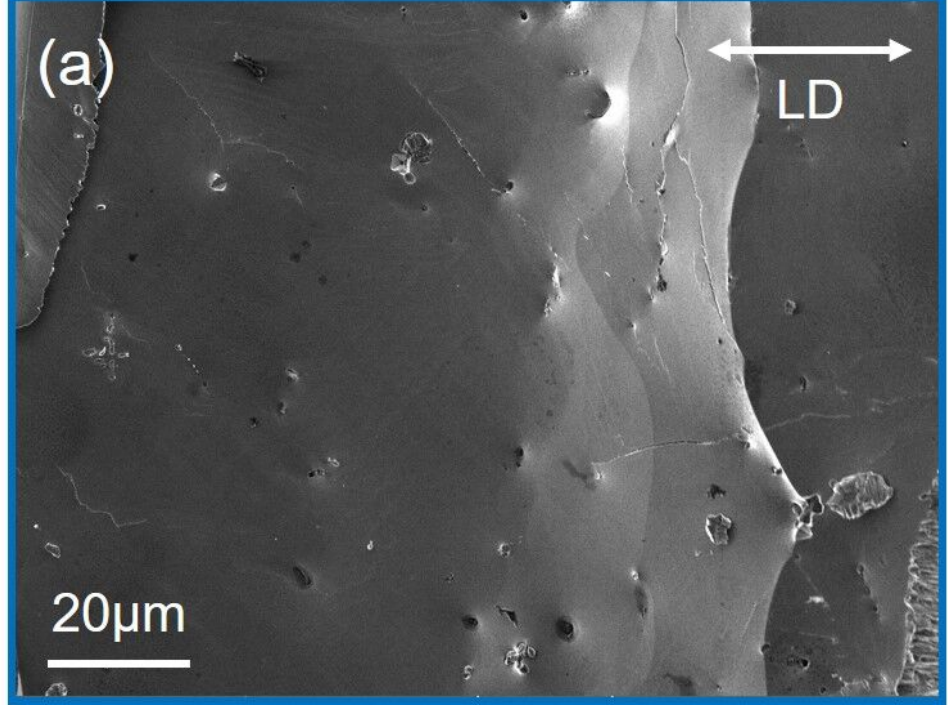

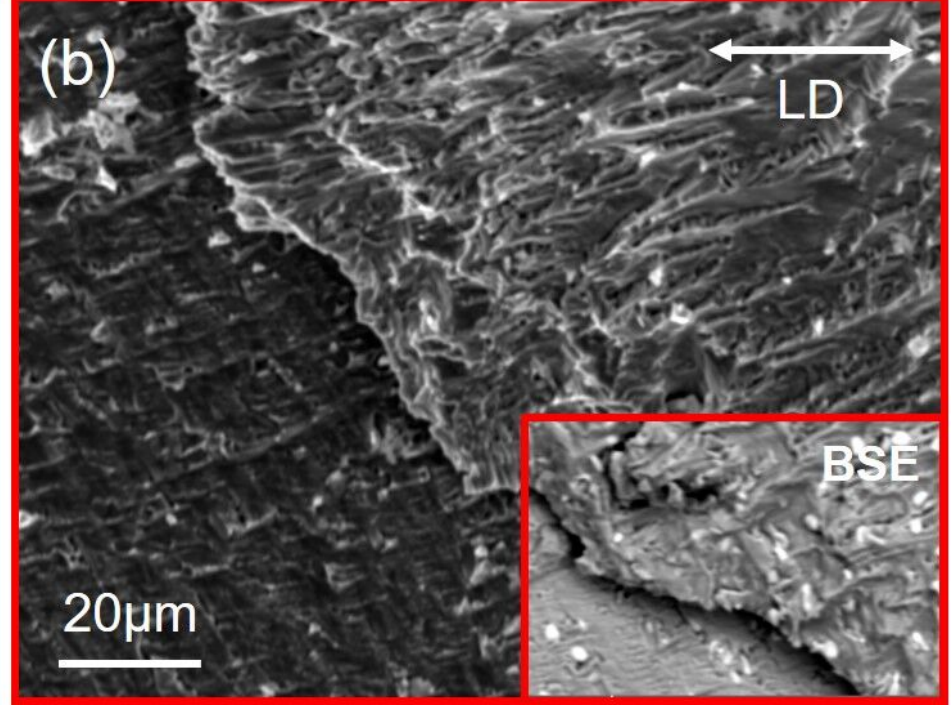

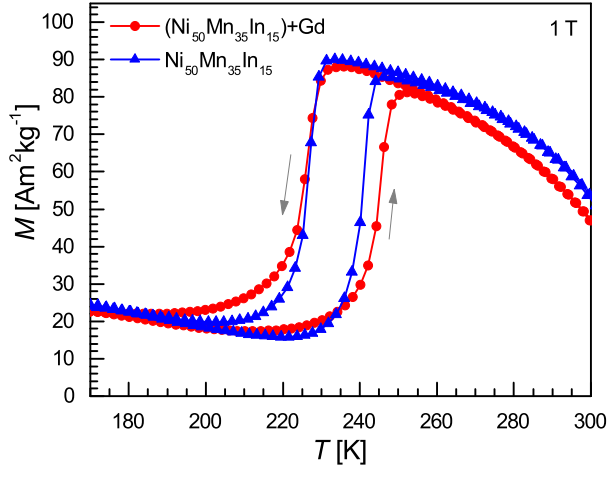

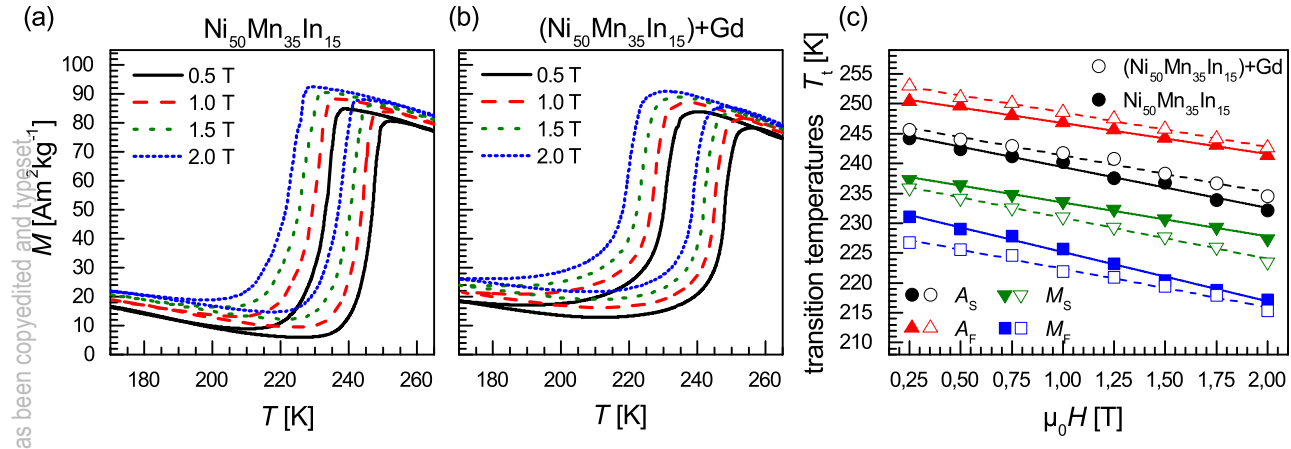

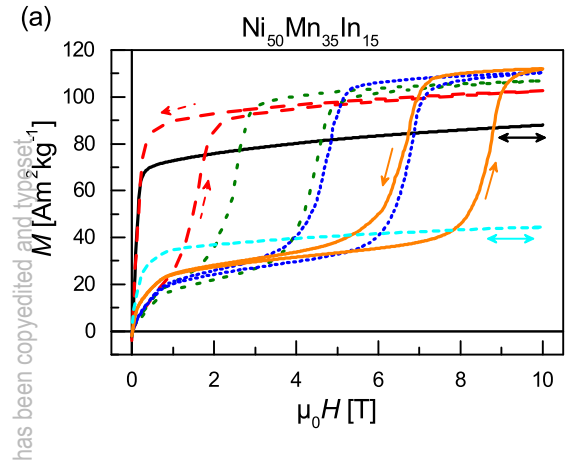

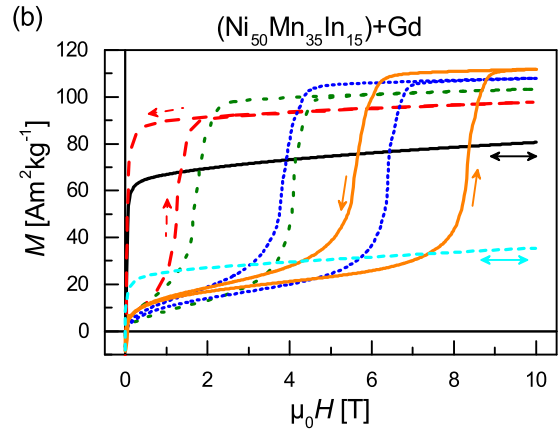

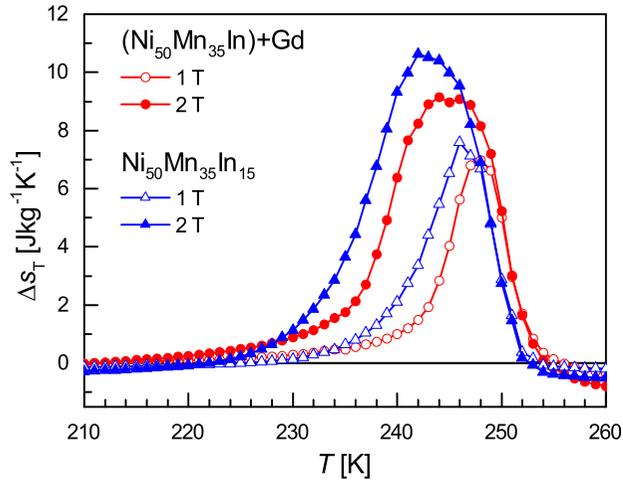

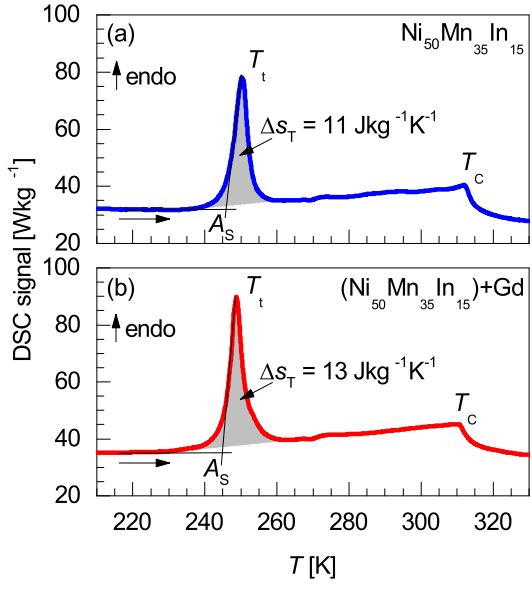

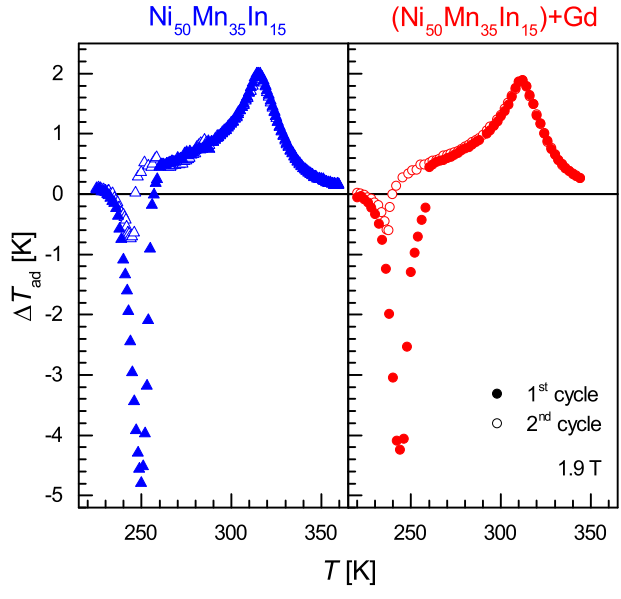